\documentclass[twocolumn,amsmath,amssymb]{revtex4}
\usepackage{graphicx}
\usepackage{dcolumn}
\usepackage{bm}
\usepackage{url}

\begin{document}

\title{Energy transfer from an individual quantum dot to a carbon nanotube}
\author{Eyal Shafran}
\author{Benjamin D. Mangum}
\author{Jordan M. Gerton}
\email{jgerton@physics.utah.edu}
\affiliation{Department of Physics and Astronomy, University of Utah, Salt Lake City, UT  84112}

\date{\today}

\begin{abstract}
A detailed understanding of energy transduction is crucial for achieving precise control of energy flow in complex, integrated systems.  
In this context, carbon nanotubes (CNTs) are intriguing model systems due to their rich, chirality-dependent electronic and optical properties.  
Here, we study the quenching of fluorescence from isolated quantum dots (QDs) upon approach of individual CNTs attached to atomic force microscope probes.  
Precision measurements of many different CNT/QD pairs reveal behavior consistent with resonant energy transfer between QD and CNT excitons via a F\"orster-like dipole-dipole coupling.  
The data reveal large variations in energy transfer length scales even though peak efficiencies are narrowly distributed around 96\%.  
This saturation of efficiency is maintained even when energy transfer must compete with elevated intrinsic non-radiative relaxation rates during QD aging.  
These observations suggest that excitons can be created at different locations along the CNT length, thereby resulting in self-limiting behavior.
\end{abstract}

\maketitle

Recently, hybrid materials composed of QDs attached to CNTs have been synthesized for a wide range of applications \cite{Biju2006, Grzelczak2006, Haremza2002, Olek2006, Li2009, Peng2009}, including photovoltaics, nanotherapeutics, bioimaging, and photocatalysis.
Each component has unique properties that make their combination highly desirable:  QDs have broad absorption spectra and size-tunable emission spectra  \cite{Norris1996}, while CNTs can be metallic with ballistic 1D charge transport, or semiconducting depending on the chiral angle of the underlying graphene lattice  \cite{Saito1998, Avouris2008}.
The interfacial area in these materials should be extremely large due to the large surface to volume ratio of both QDs and CNTs, so interactions between them are very important for their overall behavior.
In particular, the fluorescence emission from QDs is strongly suppressed when they are attached to CNTs, which indicates strong coupling between them.
Heretofore, it has not been possible to unambiguously attribute the reduced fluorescence to either charge or energy transfer between the QDs and CNTs, nor to establish limits on the coupling efficiency.  
If QD-CNT composites are indeed to be pursued for various optoelectronic applications, it is clearly important to understand the energy transduction pathways in more detail. 

It is difficult to extract a detailed understanding of the underlying energy transduction mechanisms using ensembles of QDs attached to CNTs. 
Therefore, we adopt a single-particle approach whereby we measure the interaction between single QD-CNT pairs.  
CNTs are first attached to atomic force microscope (AFM) probes via the ``pickup'' technique  \cite{Hafner2001}, and are then brought into close proximity to isolated QDs illuminated with a laser beam of well-defined polarization (Fig. 1(a)).
In a typical experiment, the CNT tip is aligned into the center of the focal spot and the sample is raster-scanned until a QD is located topographically with the AFM.
An optical image is then acquired using a unique photon counting technique   \cite{Gerton2004, Mangum2009} whereby each detected photon is correlated with the instantaneous vertical and lateral position of the CNT tip relative to the surface of the QD as the tip oscillates vertically in intermittent contact mode with a typical peak-to-peak amplitude of 50-100 nm.
A histogram of photon count rates as a function of tip height, normalized to the rate measured at the far-point of the tip oscillation, is accumulated at each lateral position, producing a 3D data set.
The 2D ($x$-$z$) fluorescence image in Fig. 1(b) demonstrates that CNTs attached to AFM probes can be used for nanometer-scale energy transfer microscopy  \cite{Mu2008};  QD-functionalized probes have been used in a similar manner previously  \cite{Ebenstein2004}.    
1D approach curves can also be extracted from the same 3D data set  \cite{Mangum2009}, or by halting the lateral scan when the CNT is centered above the QD (Fig. 2(c)).
\begin{figure}[h]
\label{fig:setup}
\centering
\includegraphics[width=3.25 in]{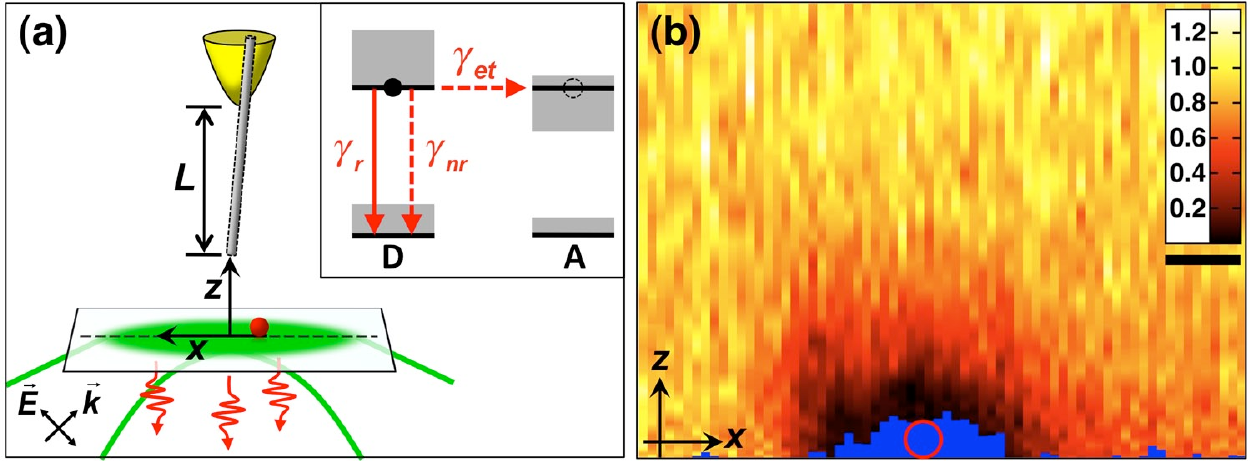}
\caption{Experimental scheme.  (a) A CNT protrudes a distance $L$ beyond a gold-coated probe.  The probe oscillates along $z$ as an isolated QD is scanned along $x$.  The sample is illuminated with an evanescent field via a focused laser beam whose wave-vector $\mathbf{k}$ is beyond the critical angle for total internal reflection.  The polarization of the evanescent wave can be parallel (as shown) or perpendicular to the CNT axis depending on the incident field direction $\mathbf{E}$.  A high numerical-aperture lens (not shown) focuses the laser beam and collects the QD emission through a glass coverslip.  The inset shows a generic level scheme for energy transfer between a donor (D) and accepter (A):  The energy transfer rate $\gamma_{et}$ competes with the intrinsic relaxation of the QD, $\gamma_0 = \gamma_r + \gamma_{nr}$.  (b) Combined topographical and optical image in an $x$-$z$ plane containing the QD.  The CNT traces out the topographical signal indicated by the blue cutout; the red circle denotes the physical size of the QD.  The normalized optical signal is given by the color scale.  The scale bar corresponds to 10 nm.}
\end{figure}

For this work, six CNTs with final protrusion lengths from $L=50$ nm to $L=165$ nm were used for $\sim$110 high-precision measurements of CdSe/ZnS QDs.
These and all measurements on an additional sample of $>$50 CNTs exhibit strong quenching of the QD fluorescence at small CNT-QD separations ($<$25 nm). 
The chiralities of the picked-up CNTs are not known, although based on the expected distribution of tubes on the growth substrates, there should be both metallic and semiconducting varieties within our sample \cite{Saito1998, Narui2009}. 
Although on very rare occasions there is some evidence of charge transfer between the QD and CNT \cite{Mu2008}, the overwhelming majority of measurements are consistent with energy transfer only (Supplementary Figure S1).   
Since energy transfer competes with the intrinsic radiative and non-radiative relaxation processes in the QD (Fig. 1(a) inset), the normalized fluorescence signal ($S$) can be generically described by:
\begin{equation}
S = \frac{Q\left(r\right)}{Q_0} = \frac{\gamma_0}{\gamma_0+\gamma_{et}\left(r\right)} = \frac{1}{1+ \gamma_{et} / \gamma_0},
\label{rate}
\end{equation}
where $Q\left(r\right)= \gamma_r/[\gamma_0+\gamma_{et}\left(r\right)]$  is the quantum yield of the QD, which depends on the position, $r$, of the CNT terminus relative to the QD surface, $\gamma_r$  is the intrinsic radiative relaxation rate of the QD, $\gamma_0 = \gamma_r + \gamma_{nr}$  is the far-field fluorescence rate, $\gamma_{nr}$  is the intrinsic non-radiative relaxation rate, $\gamma_{et}\left(r\right)$ is the position-dependent energy transfer rate, and $Q_0 = \gamma_r/\gamma_0$  is the far-field quantum yield.

The optical excitations within CNTs have been shown to be excitonic in nature \cite{Korovyanko2004, Wang2005, Ando1997, Zhao2004}, even for metallic CNTs due to reduced electron screening  \cite{Wang2007, Zeng2009}.  Thus, the energy transfer process can be described by the F\"orster theory for dipole-dipole coupling \cite{Lakowicz1999}:
\begin{equation}
\gamma_{et} / \gamma_0 = Q_0 \gamma_{et} / \gamma_r = \left(R_0/r\right)^6,
\label{forster}
\end{equation}
where $R_0$ is the so-called F\"orster radius, and $r$ is the distance between two point dipoles, the donor and acceptor.
$R_0$ depends on a number of factors, including the integrated overlap of donor-emission and acceptor-absorption spectra and the relative orientation of the donor and acceptor transition dipole moments.
The explicit dependence on  $Q_0$ given in Eqn. (\ref{forster}) implies that even for a specific CNT, the position dependence of $S$ will vary from one QD to another, and also for a particular QD during fluorescence blinking and/or oxidation-induced decay.
It is important to recognize that for every photoexcitation cycle, the QD will relax either via intrinsic processes {\em or} via energy transfer to the CNT.
Thus, the rates for intrinsic relaxation ($\gamma_0$) and energy transfer ($\gamma_{et}$) in Eqn. (\ref{forster}) are averaged over many photoexcitation/energy transfer cycles.
Furthermore, each time an energy transfer event occurs, one quantum of energy will be transferred to the CNT in the form of an exciton, which can be created anywhere along the length of the CNT.
Therefore, even when the CNT and QD are in contact, the average separation between donor and acceptor dipoles will generally be nonzero due to the physical size of the QD, and the average distance above the CNT terminus at which an exciton is generated.
For a vertically aligned CNT centered above a QD, the normalized signal should then be,
\begin{equation}
S = \left[1+\left(\frac{R_0}{z+z_0}\right)^6\right]^{-1},
\label{S}
\end{equation}
where $z$ is the vertical distance of the CNT terminus from the QD surface, and $z_0$  is the effective separation between the donor and acceptor dipoles at $z=0$.

As the photon histograms are collected, a QD will undergo a series of rapid transitions from a strongly emissive (bright) state to a weakly emissive (dark) one \cite{Nirmal1996, Gomez2006}. 
To simplify interpretation of the data, the photon signal is divided into temporal sections corresponding to bright and dark states using a simple threshold procedure, and separate approach curves are accumulated for each (Fig. 2).
The dark-state signal shows weaker fluorescence suppression since energy transfer competes less effectively with rapid internal non-radiative decay.
To avoid convoluting the analysis, only the bright-state data are compared with the modified F\"orster model.
\begin{figure}[h]
\label{fig:blink}
\centering
\includegraphics[width = 3.25 in]{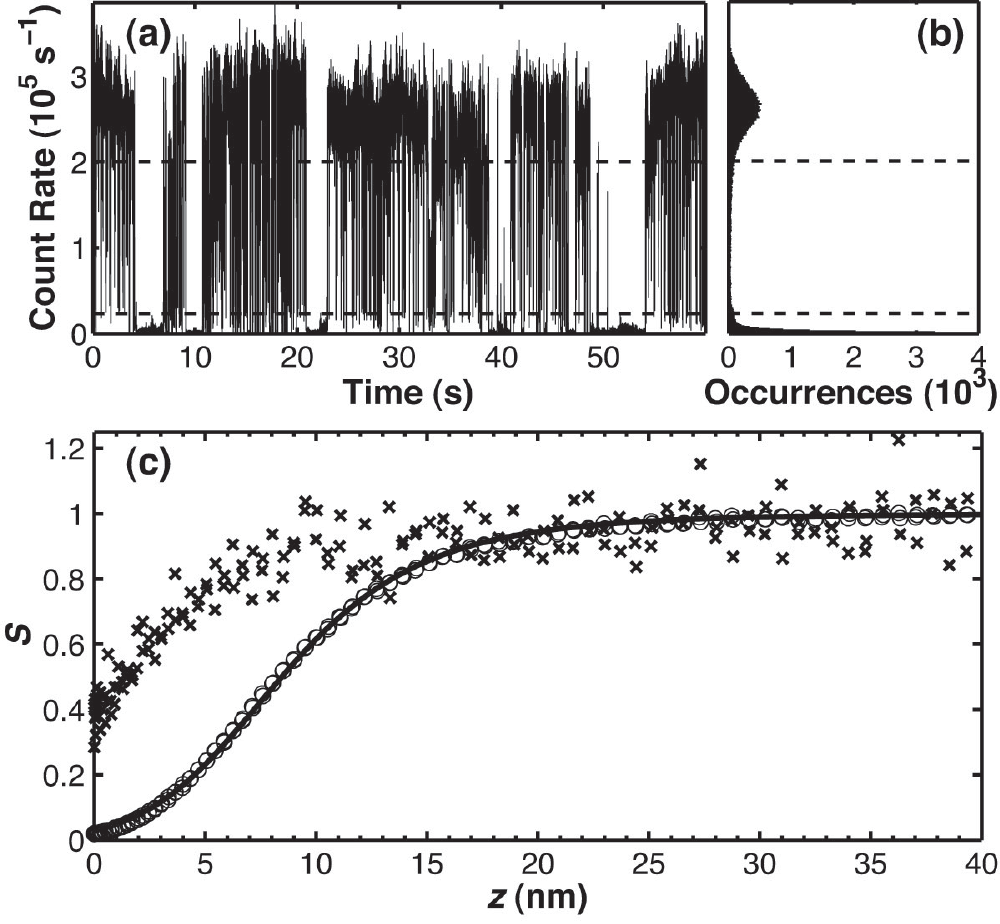}
\caption{Typical approach curves for bright and dark states of a QD.  (a) 60-second fluorescence trajectory of a QD demonstrating intermittent changes in its quantum yield.  (b) Histogram of count rates using 1 ms time bins.  The upper and lower horizontal dashed lines delineate thresholds for the bright and dark states, respectively.  (c) Vertical approach curves corresponding to the bright (open circles) and dark (cross) states.  The solid line corresponds to a fit to Eqn. (\ref{S}) with $R_0$ = 19.1 nm and  $z_0$ = 10.6 nm.}
\end{figure}

The solid curve shown in Fig. 2(c) is the best fit to Eqn. (\ref{S}) for a particular measurement.
The high quality of the fit is typical and Fig. 3(a) shows a summary of $R_0$ and $z_0$ values extracted from model fits for all six CNTs, where each $(z_0, R_0)$ pair is color coded according to the CNT length, $L$.
The fitted values for $R_0$ range from 12 nm to $\sim$40 nm, which are much larger than those for molecular fluorophores in fluorescence resonance energy transfer (FRET) experiments.
This indicates strong coupling between the QD and CNT, which requires strong absorption by the CNT at $\lambda \sim$ 600 nm \cite{Bachilo2002}, the emission wavelength for our QDs.
\begin{figure}[h]
\label{fig:length}
\centering
\includegraphics[width = 3 in]{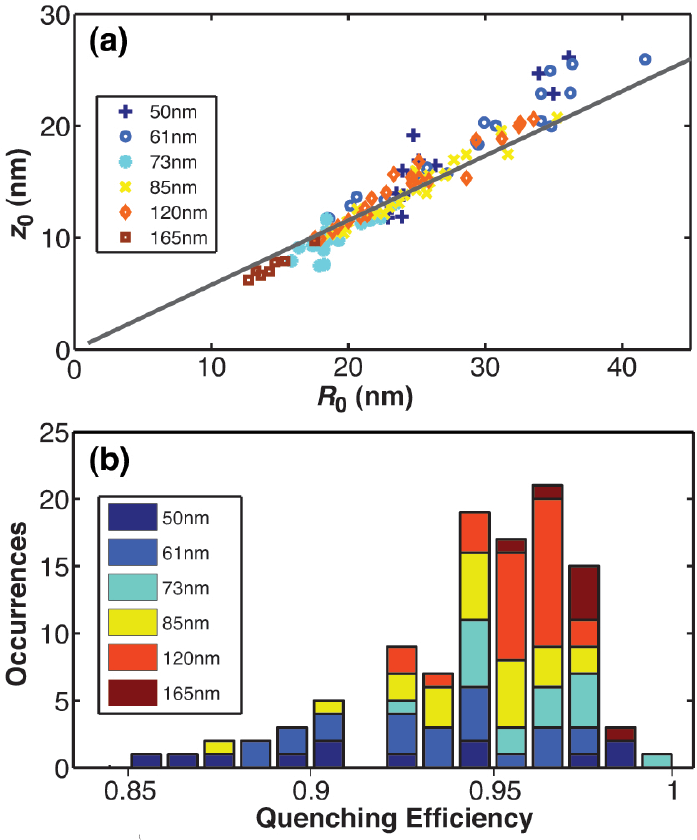}
\caption{Summary of energy transfer measurements for six CNTs of different lengths.  (a) Each point gives the value of $R_0$ and $z_0$ for a particular measurement extracted from a fit to the F\"orster model.  The points are color-coded according to CNT length.  The solid line corresponds to $R_0/\sqrt3$, as described in the text.  (b) Histogram of the measured quenching efficiencies at QD-CNT contact ($z$ = 0).}
\end{figure}

The measured correlation between $z_0$ and $R_0$ evident in Fig. 3(a) is a direct result of the 1D nature of CNTs.
In particular, the measured signal at each value of $z$ corresponds to many photoexcitation/energy transfer cycles, each of which can result in the creation of an exciton at a different position along the length of the CNT.
Stronger coupling between the QD and CNT results in a larger value of $R_0$, which increases the probability for generating an exciton further up the tube.
The values of $z_0$ extracted from the model fits are clearly much larger than the sum of the QD radius ($\sim$2 nm) and the exciton radius in the CNT (expected to be 1-4 nm depending on the CNT diameter).
Thus, although the exciton is most likely to be created near the CNT terminus where the energy transfer efficiency from the QD is largest, on average it can be created much further up the tube depending on $R_0$.

A simple estimate of the average position along the CNT at which the exciton is produced can be obtained by calculating the expectation value,
\begin{equation}
\left\langle \zeta \right\rangle = \displaystyle\frac{ \int_z^\infty \zeta \cdot E\left(\zeta,R_0\right)d\zeta}{\int_z^\infty E\left(\zeta,R_0\right)d\zeta}
\label{expect}
\end{equation}		 
where $E\left(\zeta,R_0\right) = \left[ 1+ \left(\zeta/R_0\right)^6\right]^{-1}$ is the F\"orster energy transfer efficiency between point dipoles separated by a distance $\zeta$, and $z$ and $R_0$ are as defined above.
In this context, $E$ is the probability for an energy transfer event between a donor dipole (an exciton within the QD) and an acceptor dipole (an exciton within the CNT) per photon absorbed by the QD.
Evaluating Eqn. (\ref{expect}) at $z=0$  gives $z_0 \sim \left\langle \zeta \right\rangle_{z=0} = R_0/\sqrt3$, which is plotted as the solid line in Fig. 3(a).
The strong agreement between this simple calculation and the measurements lends confidence to our interpretation of the data.

Interestingly, despite the strength of the QD-CNT coupling, the correlation between $R_0$ and $z_0$ causes the energy transfer efficiency to saturate at $z=0$, in contrast to the FRET efficiency between two point dipoles, which diverges at zero separation.
In fact, the simple analysis above predicts that the energy transfer efficiency, $1-S_{z=0}$, should saturate at a value of $(1+3^{-3})^{-1} \cong 0.96$, independent of $R_0$.
Figure 3(b) shows a stacked histogram for the {\em measured} values of the energy transfer efficiency at $z=0$ for each CNT.
There is no obvious dependence of these measurements on CNT length and importantly, the peak energy transfer efficiency for every CNT is consistent with the predicted value of 0.96.
Despite the large dynamic range in $R_0$, which reflects variations in QD-CNT coupling strength, the peak energy transfer efficiency is tightly constrained.
An important consequence of this self-limiting behavior is that the peak energy transfer efficiency should be largely independent of CNT chirality, QD-CNT spectral overlap, and the precise alignment of the QD and CNT transition dipole moments.
Since it is still very difficult to selectively grow or separate CNTs based on their chiralities, this point may be crucial in terms of using QD-CNT composites for light-harvesting applications. 

To enhance our interpretation of the measurements, a number of possible systematic effects were investigated.
First, under these conditions, gold-coated tips decrease the local illumination intensity when they are between $\sim$30 and $\sim$200 nm above the sample \cite{Mangum2009, Hdhili2003}.
To account for this, intensity profiles were measured using bare gold-coated tips, and proper normalization functions for the data were generated (Supplementary Figure S2).
This normalization procedure yields a larger uncertainty for shorter CNTs, resulting in a broader distribution of measured quenching efficiencies, as shown in Fig. 3(b).  
At very short range, a gold-coated tip can increase the local illumination intensity (for vertical polarization), and can also quench the fluorescence emission directly, but since the shortest CNT in our study is 50 nm long, these effects can be neglected.
Optical scattering from the CNT itself might also modify the illumination intensity and/or the QD radiative rate $\gamma_r$, particularly for vertically polarized illumination.
No dependence of the measurements on polarization direction is observed (Supplementary Figure S3), and indeed the data summarized in Fig. 3 contain many measurements using each polarization.
This demonstrates that any CNT-induced modification to the illumination intensity or $\gamma_r$ is either too weak or too confined near the CNT terminus to be detected at the QD core, and also that the emission dipole moment of the QD is not correlated with the excitation polarization.
Finally, it is possible for a CNT to buckle under a compressive axial load, such as applied during AFM imaging.
Indeed, buckling events lead to easily detectable asymmetry in the shape of measured approach curves, since these contain information corresponding to both the approach and retraction of the CNT during its oscillation cycle (Supplementary Figure S4), and also correlate strongly with poor AFM performance.
None of the measurements summarized in Fig. 3 exhibit the signatures of buckling.

It is important to recognize that each CNT in Fig. 3 was used to measure several individual QDs, each of which likely had a different intrinsic quantum yield, $Q_0$.
Such differences should cause $R_0$ to vary in proportion to $\sqrt[6]{Q_0}$ (Eqn. (\ref{forster})), while $R_0$ is not sensitive to disparities in the absorption cross section from one QD to another since the fluorescence data are normalized in a self-consistent manner.
In addition to variations between QDs, oxidative damage can decrease $Q_0$ for a particular QD as it ages under ambient conditions \cite{vanSark2001}.
Figure 4(a) shows a sample of four approach curves for a particular QD as it ages over a period of $\sim$20 minutes during which a constant illumination intensity was maintained.
Also shown in Fig. 4(b) are the values of $R_0$ extracted from all the measurements during this time as a function of the far-field photon count rate, $C_0$, and a solid line that is proportional to $\sqrt[6]{C_0}$.
A reduction in the absorption cross-section \cite{Kukura2009} for long aging times (i.e., small values of $C_0$) would tend to push the data to the left of this line, in agreement with the measurements.
\begin{figure}[h]
\label{fig:die}
\centering
\includegraphics[width = 3.25 in]{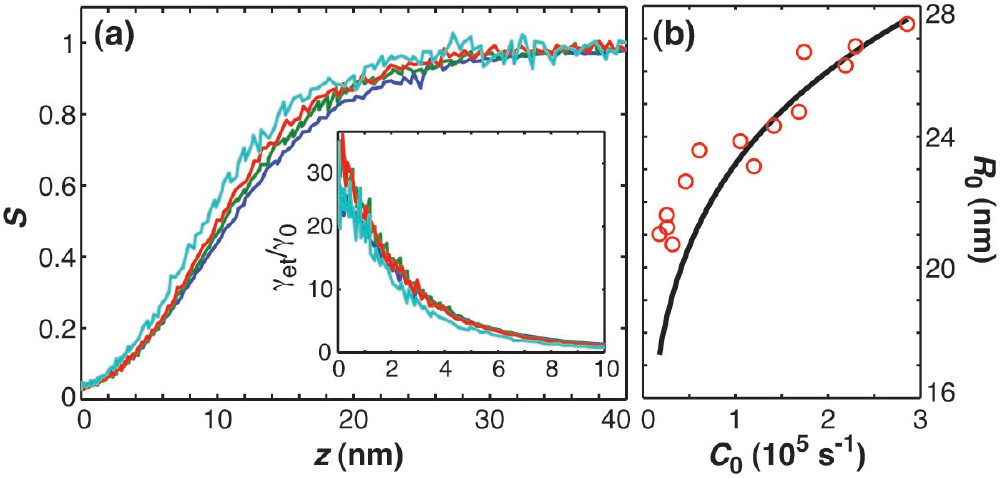}
\caption{Energy transfer measurements during QD aging.  (a) Sample of four approach curves showing a continual decrease in $R_0$ as $Q_0$ degrades.  The inset shows $\gamma_{et}/\gamma_0 = 1/S -1$ for the corresponding approach curves.  (b) Fitted values of $R_0$ as a function of the far-field photon count rate, $C_0$.  As the QD ages over a period of $\sim$20 minutes, $C_0$ decreases by a factor of $\sim$10 .}
\end{figure}

Despite the strong correlation between $R_0$ and $C_0$, the energy transfer efficiency saturates at $\sim$0.96; indeed these data are a subset of those plotted in Fig. 3.
The inset in Fig. 4(a) plots the ratio $\gamma_{et} / \gamma_0 = 1/S - 1$, which is very sensitive to minute variations when $S$ becomes small (i.e., near $z=0$). 
The remarkable consistency of the saturation value leads to the important conclusion that as $\gamma_0$ increases during aging (due to an increase in $\gamma_{nr}$), the peak energy transfer rate $\gamma_{et}(z=0)$ increases proportionally.
In FRET experiments between two point dipoles, a smaller value of $R_0$ indicates weaker coupling between the donor and acceptor, generally implying a smaller energy transfer rate.
Here, however, the one-dimensional nature of the CNT leads to a different interpretation: {\em smaller} values of $R_0$ yield {\em higher} average energy transfer rates upon repeated excitation cycles.
This occurs because when $R_0$ is small, the effective center of mass of the exciton distribution within the CNT, $\left<\zeta\right>$, is {\em closer} to the end of the CNT and thus closer to the QD, which elevates the {\em average} energy transfer rate.  
Within this context, the maximum possible energy transfer rate should occur in the limit that all acceptor excitons are created as close to the QD as possible; i.e., at the tip of the CNT.
This corresponds to a minimum dipole-dipole separation $z_{min} \geq 3$ nm: 2 nm for the radius of the QD and $\geq$1 nm for the exciton radius in the CNT.
When $R_0$ becomes smaller than about  $\sqrt3 \cdot z_{min}$, then the exciton center of mass cannot adjust toward the CNT tip any further.
Beyond this point, the energy transfer rate cannot become any larger and the energy transfer efficiency will decrease.
Evidently, the slow degradation in $Q_0$ for the QD in Fig. 4 is not sufficient to achieve these conditions and the energy transfer efficiency maintains a saturation value of $0.96$.
Dark states that occur during blinking can have sufficiently small values of $Q_0$, however, leading to reduced energy transfer efficiency (e.g., Fig. 2(c))   

In conclusion, we have made high-precision measurements of energy transfer from QDs to CNTs and have developed a simple model based on dipole-dipole coupling between excitons to explain the observed behavior.
Due to the one-dimensional nature of the CNT, the data exhibit novel features that depart from classical F\"orster theory for energy transfer between point dipoles.
In particular, we observe a strong correlation between the measured length scale ($R_0$) for efficient energy transfer and the average position of the exciton generated within the CNT ($z_0$).
This leads to a narrow distribution of the peak energy transfer efficiency and a counterintuitive increase in the energy transfer rate for smaller values of $R_0$.
Finally, both the model and measurements suggest that the peak energy transfer efficiency should be independent of CNT chirality, which has important implications with regard to the development of QD-CNT composite materials for light-harvesting applications.

\section*{Supplementary Information}
\subsection{Methods}
CNTs are grown on oxidized silicon substrates using methane-based chemical vapor deposition (CVD) and ferric nitrate catalyst nanoparticles.
The growth recipe adopted has been shown to produce mostly single-wall CNTs of both semiconducting and metallic chiralities \cite{Hafner2001, Wade2004}, although this has not been independently verified for the current work.
Following growth, CNT substrates are imaged with a commercial AFM (Asylum Research, MFP-3D) using gold-coated probes (BudgetSensors, Multi75-GB), and vertically oriented CNT whiskers can be lifted off the substrate by adhering to the sidewalls of the AFM probe.
The mechanistic details of the pickup process are not fully understood, although experimental and theoretical studies suggest that relatively large diameter CNTs (3 - 5 nm) are more likely to attach due to the increased CNT-probe interfacial area \cite{Wade2004, Solares2005}.  

Following pickup, the CNT length is measured by pressing the CNT against a smooth Si substrate while measuring the deflection of the AFM cantilever (force curve).
When the distal end of the CNT touches the substrate, the cantilever initially begins to deflect.
As more force is applied, the CNT will elastically buckle and the cantilever deflection relaxes somewhat resulting in a kink in the approach curve.
Depending on its length, a number of additional kinks are possible until finally the apex of the AFM probe comes into contact with the substrate after which a linear deflection of the cantilever is observed as the tip is further pressed into the substrate.
The measured distance between the first kink and the linear onset gives the CNT length.
When initially attached, CNTs are generally too long to possess sufficient axial stiffness for use in AFM imaging and thus they must be shortened to $<$200 nm.  This is achieved by application of short ($\sim$10 $\mu$s) voltage pulses of 10 V amplitude between the AFM probe and a doped Si substrate.
These pulses induce electrochemical etching of the distal end of the CNT, which leads to shortening in quasi-controllable steps of 10-15 nm and removes any fullerene or catalyst cap.
In these experiments, no attempt was made to successively shorten a particular CNT between fluorescence measurements.
Rather, each CNT was used for a series of measurements until it was irreversibly damaged or lost.
The lengths of the CNTs utilized extended from 50 to 165 nm, which is approximately the usable range for these types of measurements.

The experimental setup is described in more detail elsewhere \cite{Mangum2009}.
Briefly, the AFM is coupled to a homebuilt inverted optical microscope, which features a 1.4 NA oil-immersion objective, a single-photon counting avalanche photodiode (APD) module (Perkin Elmer, SPCM-AQR14), and a green He-Ne laser ($\lambda = 543$ nm).
A key element in the setup is a laser beam mask whose transmission profile is a $60^{\circ}$ annular section.
The inner diameter of the annular section blocks all subcritical rays, resulting in a purely evanescent field at the sample interface within an elongated (1.5 $\times$ 0.5 $\mu$m$^2$) focal area.
By rotating the linear polarization of the laser beam, the polarization of the evanescent field can be adjusted to be either vertical (i.e., along the CNT axis) or horizontal (i.e., in the plane of the sample).
The QDs employed are nominally 4 nm in diameter and 9 nm long and have an emission maximum at a wavelength of $\sim$605 nm (Invitrogen, QDOT605 ITK).
After diluting in toluene to a concentration of $\sim$10$^{-9}$ M, 50 $\mu$l of the QD solution was then pipetted onto a cleaned and etched glass coverslip and left to dry in a laminar flow hood.
The final surface density of QDs on the coverslip was $<$0.1 $\mu$m$^{-2}$. 
With such a low QD surface density, it is easy to locate isolated QDs using the AFM topography.

\makeatletter
\renewcommand{\thefigure}{S\@arabic\c@figure}
\renewcommand{\theequation}{S\@arabic\c@equation}
\renewcommand{\thesection}{S\@arabic\c@section}
\makeatother 
\setcounter{figure}{0}
\setcounter{equation}{0}
\setcounter{section}{0}

\subsection{Evidence against charge transfer}
When a QD becomes charged, its quantum yield is reduced significantly because subsequently excited excitons can be non-radiatively quenched via an Auger recombination process \cite{Gomez2006}.
This is, in fact, related to the probable mechanism for QD blinking \cite{Nirmal1996}:  an exciton's electron or hole can become trapped in a localized defect state (e.g., dangling bonds) at the surface, leaving a residual charge in the core \cite{Gomez2006}.
In blinking, the quantum yield will remain low (off) until the trapped charge recombines in the core, upon which it returns to a high (on) value.
Histograms of the persistence times for both the "on" and "off" states will typically exhibit a power-law dependence on time, with an exponent of $\sim-1.5$.  
Thus, if a charge transfer event between the QD and an approaching CNT occurred, the blinking statistics should be altered inasmuch as the neutralization time is slower than the bin time (inverse sample rate).
Figure S1 shows a comparison between the on and off persistence times for a particular QD when a CNT is oscillating directly above it during an energy transfer measurement, and when the CNT is not present.
To facilitate comparison of the histograms, the CNT data has been multiplied by an appropriate factor, which vertically shifts the data on the log-log scale:  no discernible difference between the power-law distributions is observed.
The tip oscillation period in this case is $\sim$15 $\mu$s, so there are many tip oscillations per 1 ms time bin.
Thus, there is no evidence for charge transfer in these measurements, assuming the neutralization time is longer than the 1 ms.
\begin{figure}[h]
	\centering
		\includegraphics[width=3.25 in]{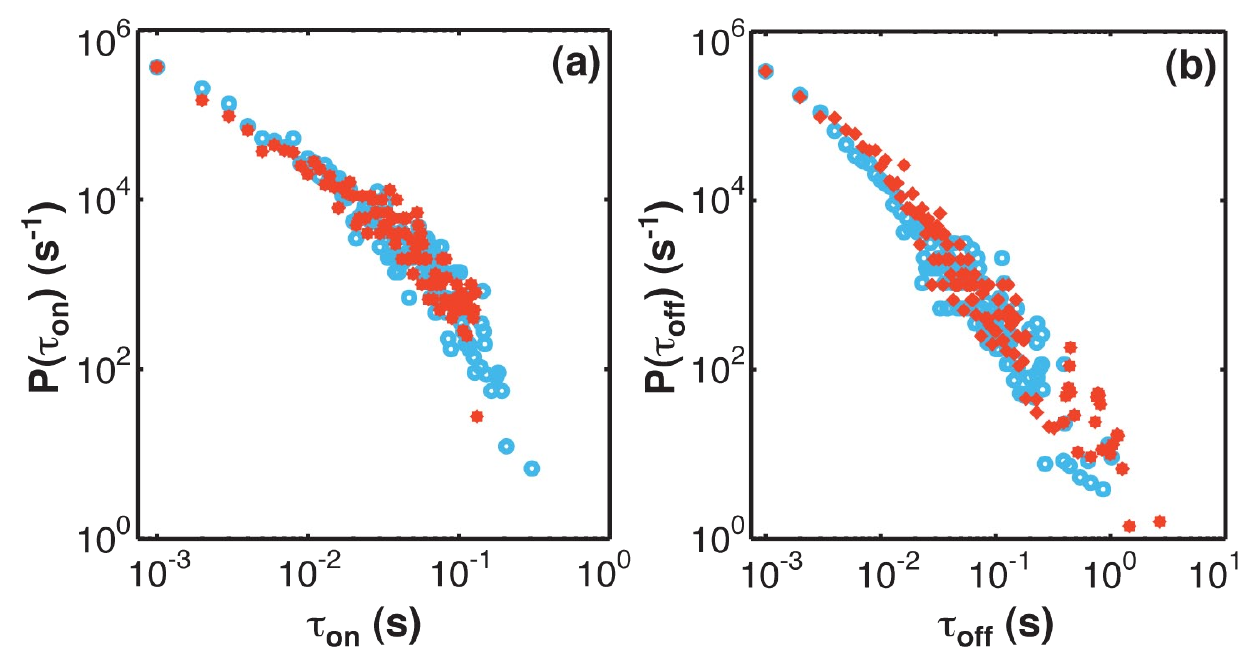}
	\label{fig:onoff}
	\caption{Comparison of quantum dot blinking statistics with and without a carbon nanotube.  (a) Histogram of persistence times for the "on" state.  (b) Histogram of persistence times for the "off" state.  In both panels, the open blue circles represent measurements made with the CNT oscillating above the QD during acquisition of an approach curve and the solid red circles are without the CNT present.  The vertical axis is the "on" or "off" persistence probability density \cite{Peterson2009}.}
\end{figure}
 
The symmetric shape of the approach curves provides additional evidence against charge transfer.
As described in the text and in previous work \cite{Gerton2004, Mangum2009}, these approach curves are acquired while the CNT-tip oscillates above the QD with a frequency of $\sim$70 kHz.
The detected photons are time-tagged and initially correlated with the instantaneous tip oscillation phase, and these phases are subsequently mapped to the instantaneous height of the tip using the calibrated tip oscillation amplitude.
Thus, the measured approach curves contain information on both the approach and retraction of the CNT tip towards and away from the sample surface.
If a charge transfer event occurred as a CNT approached the QD, the QD would go dark for some period until the QD charge was neutralized, leading to an asymmetric shape of the approach curve.
Note that large asymmetries in some approach curves are sometimes observed (e.g., see Fig. S4 below), but we interpret these as resulting from CNT buckling.
Either way, none of the data analyzed exhibited such asymmetry, so we conclude that there is no charge transfer if the neutralization time is longer than the tip-oscillation period, or $\sim$8 $\mu$s.
Since our QDs sit atop a non-conductive glass substrate, and since the experiments are performed in a relatively low humidity environment ($<$10$\%$), neutralization times shorter than this are not expected.

\subsection{Normalization procedure}
The modified F\"orster model developed in the text predicts a signal of the form:
\begin{equation}
S = \left[1+ \left[R_0 / \left(z+z_0\right)\right]^6\right]^{-1}.
\label{S2}
\end{equation}
Equation (\ref{S2}) assumes that the fluorescence signal is properly normalized to the far-field value (i.e., $z \rightarrow \infty$), and that the detected signal is due purely to an interaction between the QD and CNT.
However, it has been shown that metal tips can also quench fluorescence at short distances and can also modify the local optical intensity at the QD at distances up to $\sim$200 nm \cite{Mangum2009}.
Since the shortest CNT used in our study was 50 nm long, we are only interested the long-range effects and can neglect quenching.
In particular, as it approaches the QD, the gold tip to which a CNT is attached suppresses the fluorescence signal, as seen in Fig. S2.
Due to the long length scale for this effect, we believe it results from a reduction in the local intensity at the QD caused by interference between the excitation light and the light scattered off the tip, rather than from energy transfer from the QD to the gold tip.
The scattering from the tip is expected to be stronger for vertically polarized light compared to horizontal polarization, which leads to the longer decay length seen in Fig. S2.
\begin{figure}[h]
	\centering
		\includegraphics[width = 3 in]{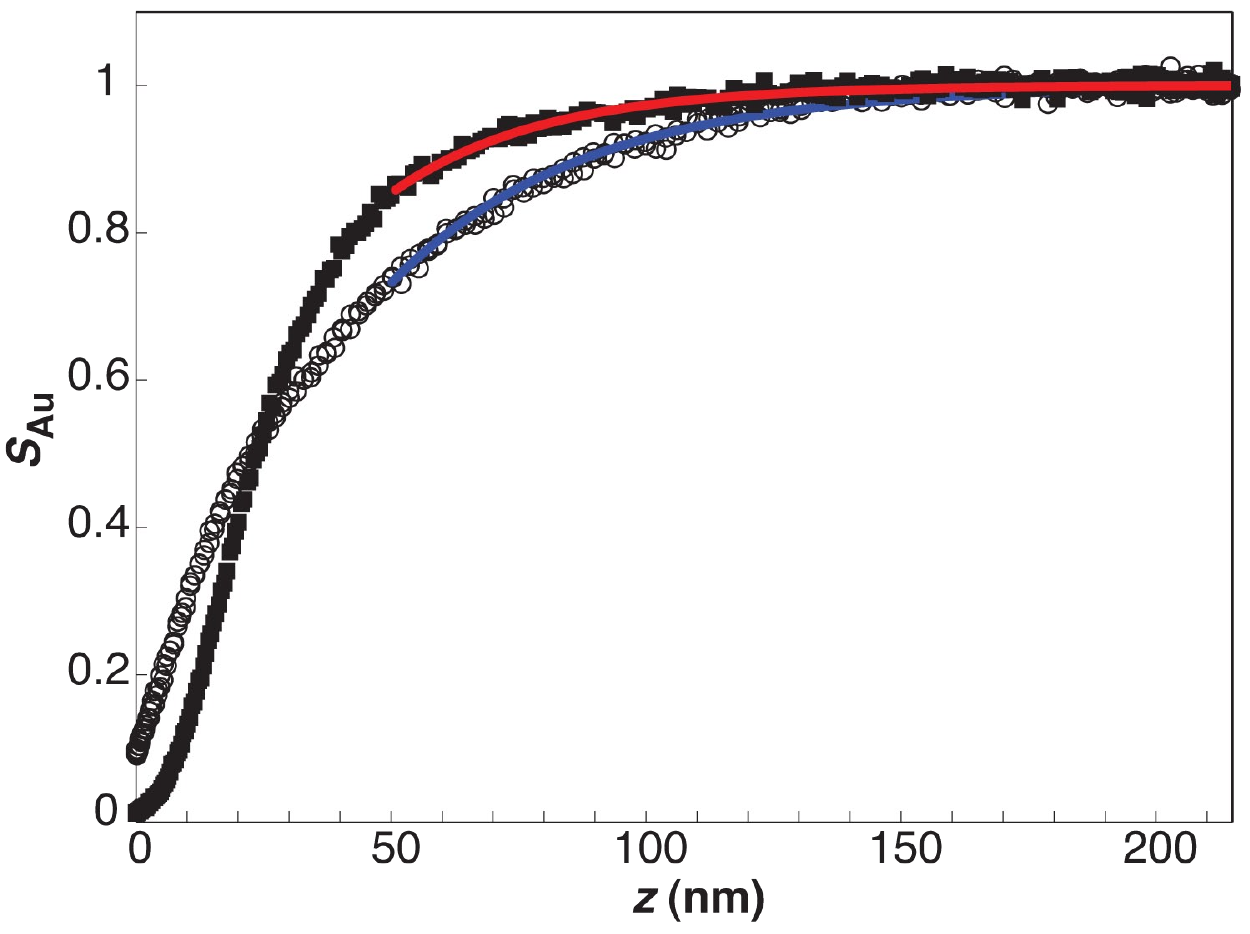}
	\label{fig:Gold}
	\caption{Approach curves for a bare gold tip. The open circles correspond to illumination with vertical polarization and the solid squares are for horizontal polarization. The solid lines are fits to an exponential decay function of the form, $S_{Au}(z) = 1- A\exp(z/z_d)$ in the range $z>50$ nm.}
\end{figure}

The approach curves in Fig. S2 for both vertical and horizontal polarization are described well by an exponential decay function, $S_{Au}(z) = 1- A\exp(z/z_d)$, for $z > 50$ nm.
Thus, to account for the gold-induced intensity suppression, the approach-curve data for CNT tips was first fit to $S_{Au}(z+L)$, where $L$ is the measured length of the specific CNT, and the parameters $A$ and $z_d$ are extracted from the fit. 
Only data in the range $z>25$ nm are used in the fit, since in this range the contribution from the CNT is smaller than that from the gold.
The measured signal is then normalized by the contribution of the gold for the entire range of the measurement: $S(z) = S_{CNT}(z)/S_{Au}(z+L)$.

\subsection{Dependence on illumination polarization}
There are two primary possibilities that could lead to a polarization dependence in the CNT approach curve measurements.
First, direct scattering of the excitation light by the CNT should lead to field enhancement at the CNT terminus via the lightning-rod effect when the polarization is parallel to the CNT axis.
For transverse polarization, no field enhancement should result and in fact a reduction in excitation intensity might occur.
Secondly, if the emission dipole orientation of the QD is correlated with its absorption dipole, then $R_0$ should depend on the excitation polarization direction via the dipole-dipole orientation overlap factor.
The data presented in the manuscript was composed of many measurements with both axial and transverse polarization directions and no systematic difference was observed.
To demonstrate this, Fig. S3 shows the measured values of peak quenching efficiency (at $z=0$) and the quenching decay length for each CNT and both polarization directions.
The decay length is the measured tip-sample separation at half the peak quenching efficiency.
\begin{figure}[h]
	\centering
		\includegraphics[width = 3 in]{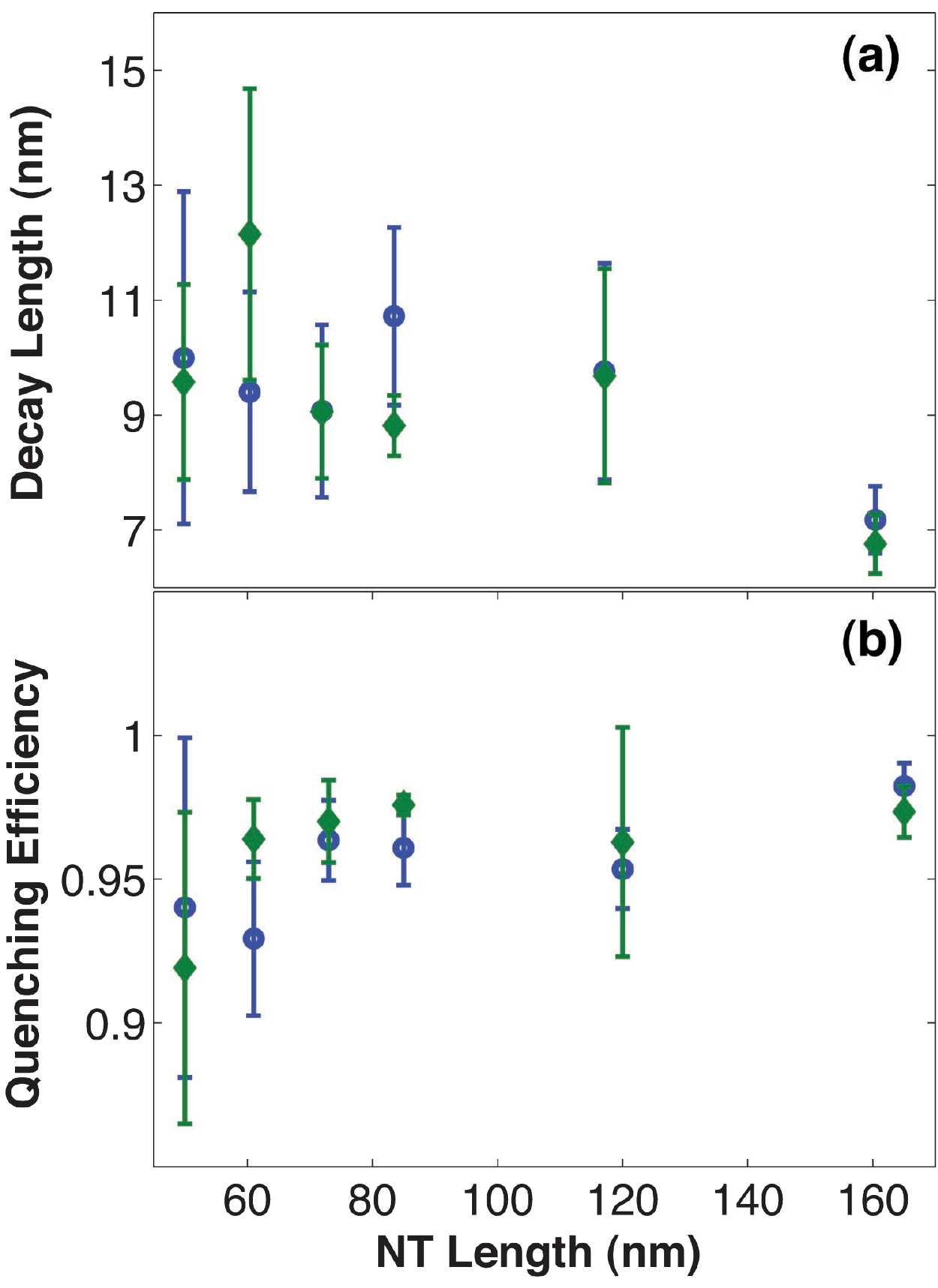}
	\label{fig:pol}
	\caption{Polarization dependence of energy transfer measurements.  (a) Decay length, and (b) peak quenching efficiency for both axial (green diamonds) and transverse (blue circles) excitation polarization.  The error bars represent the standard deviation of the measurements for each CNT length.}
\end{figure}

\subsection{Nanotube buckling}
Occasionally, we observe obvious asymmetry in the shape of an approach curve, as shown in Fig. S4.
In addition, asymmetric approach curves correlate strongly with poor AFM performance across the entire sample, not just above a QD.
Thus, we interpret such data as resulting from buckling of the CNT while the cantilever oscillates during intermittent contact mode AFM imaging.
It is important to recall that all the measurements are obtained in this mode and that the approach curves are acquired by correlating the arrival times of the detected photons with the instantaneous phase of the tip oscillation.
Thus, if the CNT buckles as the cantilever swings through its downward trajectory (i.e. toward the sample surface), then upon retraction, it will elastically unbuckle at some point and lift away from the surface.
As is often observed for normal AFM tips, the CNT is likely to adhere to the surface upon retraction, until the cantilever spring constant overwhelms the adhesion forces, at which point the CNT will release from the surface.
This causes an asymmetry in the phase histogram and associated approach curve:  the approach side of the curve exhibits a smooth decay in the fluorescence with a shape characteristic of a normal (non-buckled) measurement, while the retraction side of the curve is more steep.
In addition, there is a flat bottom to the approach curve corresponding to the duration of time during which the CNT is in a buckled state, and apparently some portion of the CNT is in direct contact with the QD.
Occasionally the flat portion of the approach curve exhibits more complicated features, which presumably correspond to the CNT tip sliding away from the QD location.
In addition, we observe a higher rate of buckling with larger tip-oscillation amplitudes and harder tapping (lower set point for the oscillation amplitude relative to the free amplitude), and a corresponding increase in the width of the flat portion of the approach curve.
\begin{figure}[t]
	\centering
		\includegraphics[width = 3.25 in]{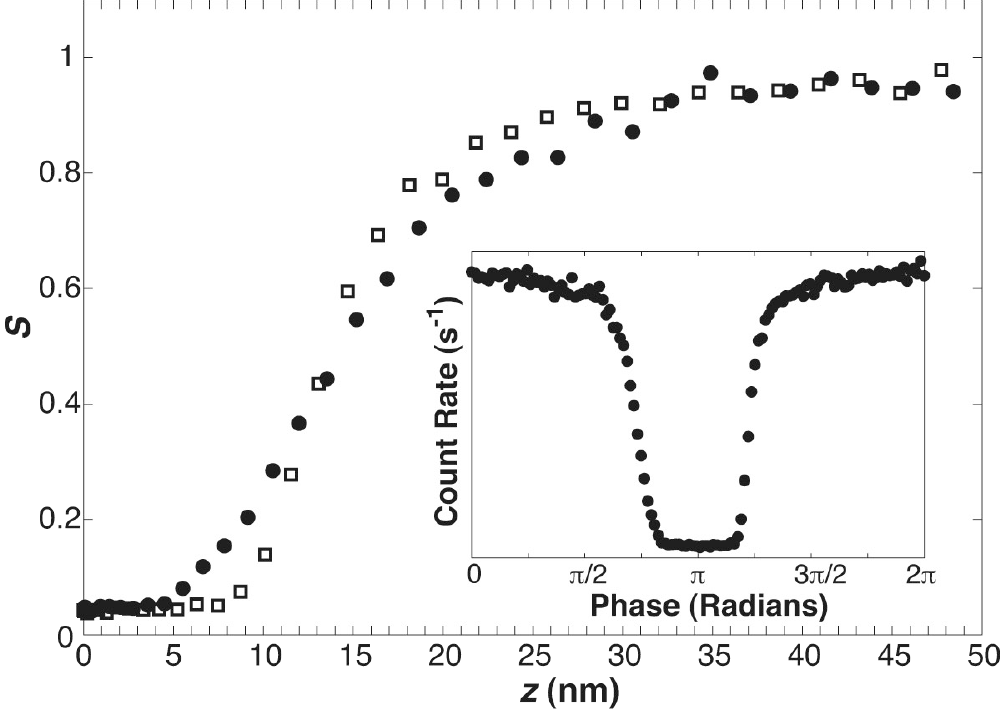}
	\label{fig:buckle}
	\caption{Buckling of a CNT during an approach curve measurement.  The closed circles correspond to approach of the CNT toward the QD, while the open squares correspond to its retraction.  The precise location of QD-CNT contact is difficult to determine in this case.  The inset is the phase histogram generated during the measurement from which the approach curves are extracted.  This phase histogram is generated by correlating the arrival times of detected photons with the instantaneous phase of the tip oscillation.}
\end{figure}

\begin{acknowledgments}
The authors gratefully acknowledge helpful discussions with Eugene Mishchenko and Dane McCamey.  This work was supported by a Cottrell Scholar Award from the Research Corporation for Science Advancement and an NSF CAREER Award number DBI-0845193.
\end{acknowledgments}


\end{document}